\documentclass[12pt]{iopart}
\usepackage{graphicx}

\renewcommand{\d}{\mathrm{d}}
\renewcommand{\pt}{p_\mathrm{T}}
\newcommand{\mt}{m_\mathrm{T}}
\begin{document}

\title{Low Mass Dimuon Production in p-A Collisions at $\sqrt{s} = 27.5$~GeV with NA60}

\author{A.~Uras\footnote[1]{E-mail: antonio.uras@cern.ch} for the NA60 Collaboration:\\R.~Arnaldi$^{1}$,  K.~Banicz$^{2,3}$,  J.~Castor$^{4}$, B.~Chaurand$^{5}$, W.~Chen$^{6}$,  C.~Cical\`o$^{7}$,   A.~Colla$^{8}$, P.~Cortese$^{8}$, S.~Damjanovic$^{2,3}$,  A.~David$^{2,9}$, A.~de~Falco$^{10}$,  A.~Devaux$^{4}$,  L.~Ducroux$^{11}$, H.~En'yo$^{12}$, J.~Fargeix$^{4}$, A.~Ferretti$^{8}$,  M.~Floris$^{10}$, A.~F\"orster$^{2}$, P.~Force$^{4}$, N.~Guettet$^{2,4}$, A.~Guichard$^{11}$,  H.~Gulkanian$^{13}$,  J.~M.~Heuser$^{12}$, M.~Keil$^{2,9}$,  Z.~Li$^{6}$, C.~Louren\c{c}o$^{2}$, J.~Lozano$^{9}$, F.~Manso$^{4}$,  P.~Martins$^{2,9}$,   A.~Masoni$^{7}$, A.~Neves$^{9}$, H.~Ohnishi$^{12}$,  C.~Oppedisano$^{1}$, P.~Parracho$^{9}$, P.~Pillot$^{11}$,  T.~Poghosyan$^{13}$, G.~Puddu$^{10}$, E.~Radermacher$^{2}$, P.~Ramalhete$^{2,9}$,  P.~Rosinsky$^{2}$, E.~Scomparin$^{1}$, J.~Seixas$^{9}$,  S.~Serci$^{10}$, R.~Shahoyan$^{2,9}$,  P.~Sonderegger$^{9}$, H.~J.~Specht$^{3}$, R.~Tieulent$^{11}$,  A.~Uras$^{10,11}$,  G.~Usai$^{10}$,  R.~Veenhof$^{9}$, H.~K.~W\"ohri$^{9}$}

\address{{$^{~1}$ INFN, Sezione di Torino, Italy}}
\address{{$^{~2}$ CERN, 1211 Geneva 23, Switzerland}}
\address{{$^{~3}$ Physikalisches~Institut~der~Universit\"{a}t Heidelberg,~Germany}}
\address{{$^{~4}$ Universit\'e Blaise Pascal and CNRS-IN2P3, Clermont-Ferrand, France}}
\address{{$^{~5}$ LLR, Ecole Polytechnique and CNRS-IN2P3, Palaiseau, France}}
\address{{$^{~6}$ BNL, Upton, NY, USA}}
\address{{$^{~7}$ INFN, Sezione di Cagliari, Italy}}
\address{{$^{~8}$ Dip.~di Fisica Sperimentale dell'Universit\`a di Torino and INFN, Torino,~Italy}}
\address{{$^{~9}$ Instituto Superior T\'ecnico, Lisbon, Portugal}}
\address{{$^{~10}$ Dipartimento di Fisica dell'Universit\`a di Cagliari and INFN, Cagliari, Italy}}
\address{{$^{~11}$ IPNL, Universit\'e Claude Bernard Lyon-I and CNRS-IN2P3, Villeurbanne, France}}
\address{{$^{~12}$ RIKEN, Wako, Saitama, Japan}}
\address{{$^{~13}$ YerPhI, Yerevan Physics Institute, Yerevan, Armenia}} 

%\ead{antonio.uras@cern.ch}

\begin{abstract}
The NA60 experiment has studied low-mass muon pair production in proton-nucleus collisions with a system of Be, Cu, In, W, Pb and U targets using a 400 GeV/$c$ proton beam at the CERN SPS. The mass spectrum is well described by the superposition of the two-body and Dalitz decays of the light neutral mesons $\eta$, $\rho$, $\omega$, $\eta'$ and $\phi$. A new high-precision measurement of the electromagnetic transition form factors of the $\eta$ and $\omega$ mesons is presented, complemented with a measurement of the temperature parameter of the $\rho$ meson in cold nuclear matter. The $\pt$ spectra for the $\omega$ and $\phi$ mesons are extracted in the full $\pt$ range accessible, up to $\pt = 2$~GeV/$c$. The nuclear dependence of the production cross sections for the $\eta$, $\omega$ and $\phi$ mesons has been investigated in terms of the power law $\sigma_\mathrm{pA} \propto \mathrm{A}^\alpha$, and the $\alpha$ parameter was studied as a function of $\pt$.

\end{abstract}

%\section{Introduction} 
\noindent The study of the production of low mass vector and pseudoscalar mesons in p-A collisions represents a natural baseline for the heavy-ion observations, providing a reference for different observable in cold nuclear matter, as
strangeness production as a function of~A. In addition, the study of the nuclear dependence of particle properties in p-A, as the $\pt$ spectra and the production cross sections, is an effective tool to understand the dynamics of soft hadron interactions. In order to address these items, the NA60 experiment complemented the In-In data with a high statistics dimuon sample collected in p-A, exposing to the beam six target materials: Be, Cu, In, W, Pb and~U. In this way, a comprehensive and detailed study of the production of the $\eta$, $\rho$, $\omega$ and $\phi$ mesons as a function of~A and $\pt$ has been performed for the first time.

%\section{Apparatus and Event Selection}
A description of the NA60 apparatus can be found for example in~\cite{apparatus}. The produced dimuons are identified and measured by the muon spectrometer placed after a hadron absorber. The latter allows one to select the highly rare dimuon events but induces at the same time multiple scattering and energy loss on the muons, thus degrading the mass resolution of the measurement made in the spectrometer. To overcome this problem, NA60 already measures the muons before the absorber, with a silicon pixel spectrometer. The final sample of dimuons retained for the analysis is defined according to the quality of the matching between the reconstructed tracks in the two spectrometers: this leads to a total sample of $\sim180\,000$ dimuons. When the identification of the production target is requested, however, stricter cuts must be applied, reducing the total sample to $\sim80\,000$ dimuons. The small component of the combinatorial background ($\pi$ and $K$ decays) is estimated via an event mixing technique and subtracted from the real data, with an overall precision of a few percent. More details can be found in~\cite{HP2010}.

\begin{figure}[htbp]
  \vspace{-0.5cm}
  \includegraphics[width=.40\textwidth]{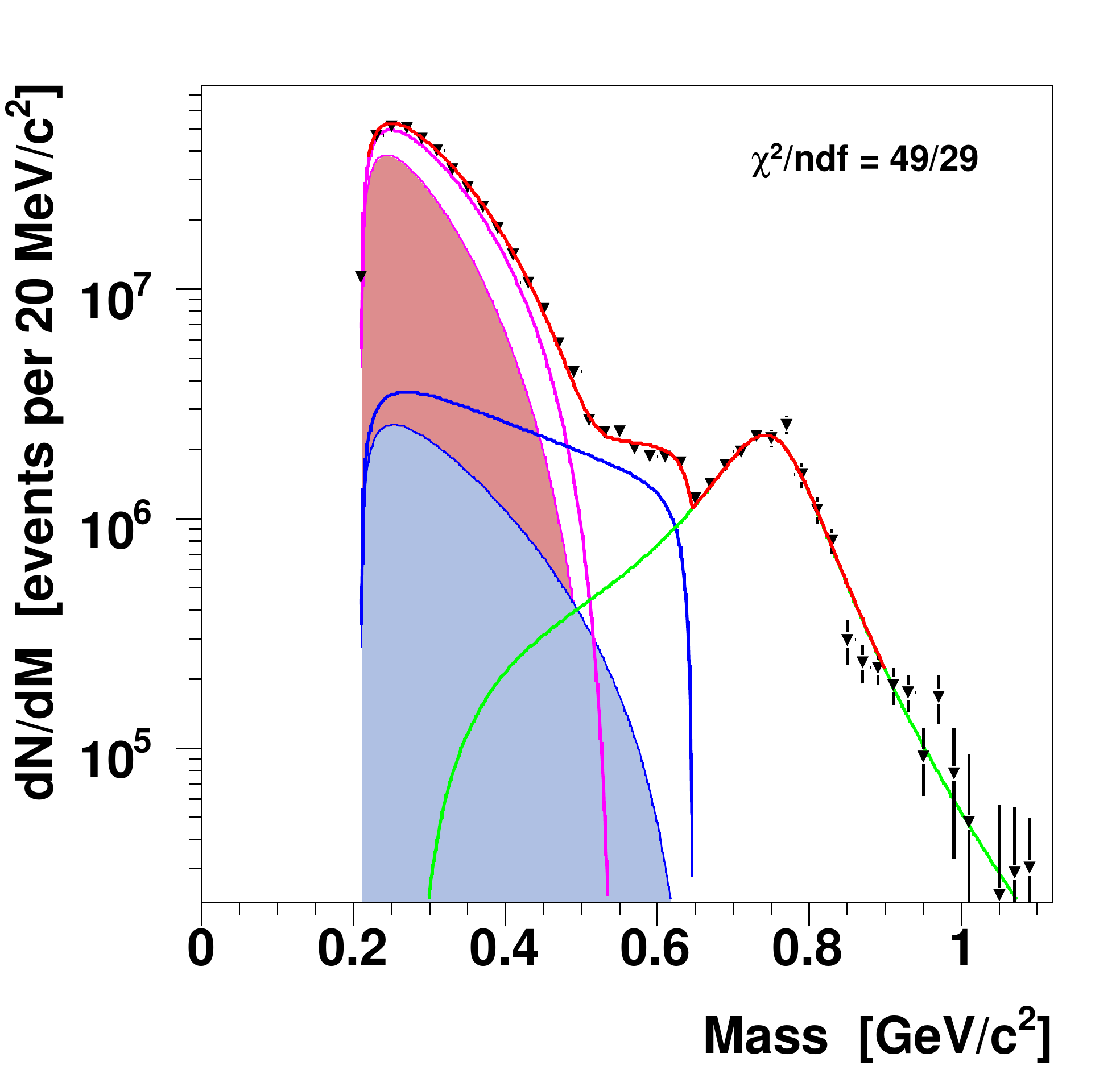} \hspace{-.15\textwidth}
  \begin{minipage}[b]{.62\textwidth}
  \caption[\textwidth]{\label{massSpectrumFit} Fit on the acceptance corrected mass spectrum, with the contribution of the $\eta\to\mu\mu\gamma$, $\omega\to\mu\mu\pi^0$ and $\rho\to\mu\mu$ processes. Details can be found in~\cite{Uras:HQ2010}. \vspace{0.8cm}}
  \end{minipage}
\vspace{-0.2cm}
\end{figure}

%\section{Target-Independent Observables}
The mass spectrum is well described by the superposition of the two-body and Dalitz decays of the light neutral mesons $\eta$, $\rho$, $\omega$, $\eta'$ and $\phi$. Considering the target-integrated data sample, the electromagnetic transition form factors of the $\eta$ and $\omega$ mesons have been studied through the Dalitz decays $\eta \to \mu^+ \mu^- \gamma$ and $\omega \to \mu^+ \mu^- \pi^0$. These form factors are usually expressed in terms of the pole parametrization $|F|^2 = (1 - M^2/\Lambda^2)^{-2}$, with the Vector Meson Dominance (VMD) model giving theoretical predictions both for $\Lambda_{\eta}^{-2}$ and $\Lambda_{\omega}^{-2}$, which have now been tested with the NA60 p-A data. While the details of the analysis can be found elsewhere~\cite{Uras:HQ2010}, we report here the final values, extracted from the fit on the corrected mass spectrum shown in \figurename~\ref{massSpectrumFit}, currently the most precise measurements available for the $\mathrm{\Lambda}^2$: $\mathrm{\Lambda}_\eta^{-2} = 1.950\ \pm\ 0.059$~(stat.) $\pm\ 0.042$~(syst.) and $\mathrm{\Lambda}_\omega^{-2} = 2.241\ \pm\ 0.025$~(stat.) $\pm\ 0.028$~(syst.), both expressed in (GeV/$c^2$)$^{-2}$. They are in perfect agreement with the values obtained by the analysis on the NA60 peripheral In-In data~\cite{na60} and the Lepton-G results~\cite{landsberg}, confirming the agreement with the VMD model for the $\eta$ meson and the discrepancy in the case of the $\omega$ meson. An improved description of the $\omega$ form factor, given by an alternative theoretical approach recently appeared in~\cite{Leupold}, still underestimates the part close to the kinematical cut-off. Finally, it was possible to isolate and study the line shape of the $\rho$ meson, which was parametrized with the function 
$\frac{\d N}{\d M} \propto \frac{\sqrt{1-4m^2_\mu/M^2}\left( 1+2m^2_\mu/M^2\right)\left( 1-4m^2_\pi/M^2\right) ^{3/2}}
{\left( m^2_{\rho}-M^2\right)^2+m^2_\rho\Gamma^2_\rho(M)} \left(MT\right)^{3/2} e^{-\frac{M}{T}}$. This parametrization is well suited to describe the $\rho$ line shape in cold nuclear matter, with the Boltzmann-like factor $\left(MT\right)^{3/2} e^{-\frac{M}{T}}$ playing a crucial role in shaping the low mass tail. The $T$ parameter is found to be $161\pm7$~(stat.) $\pm7$~(syst.)~MeV. This is the first measurement of the effective temperature of the $\rho$ meson in cold nuclear matter: it is in agreement with the measurement in peripheral In-In~\cite{na60}, and consistent with the value of 170~MeV obtained by statistical model fits on particle ratios in p-p interactions.

\begin{figure}[htbp] 
  \begin{center}
    \hspace*{-0.04\textwidth}\includegraphics[width=0.26\textwidth]{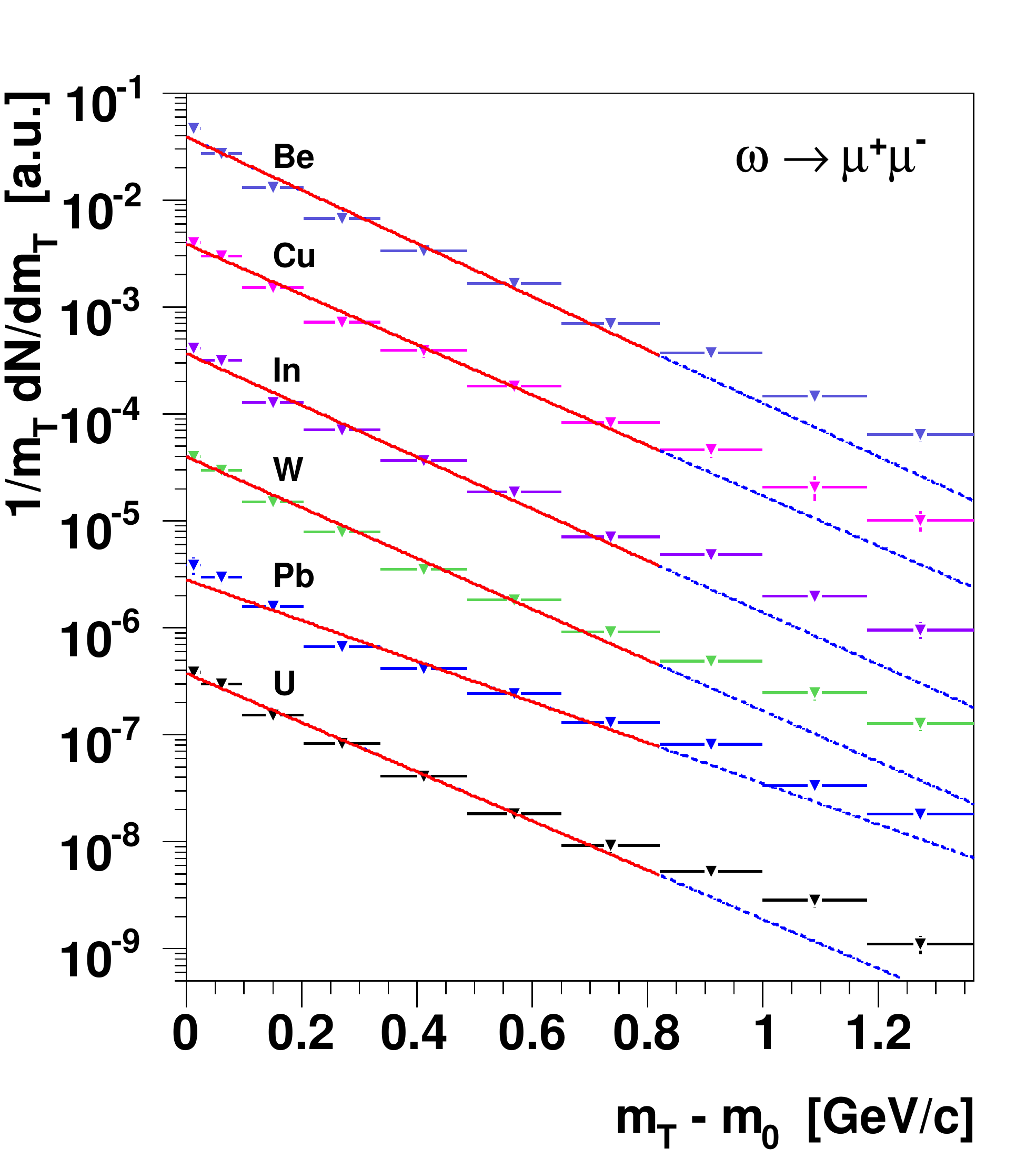}\hspace*{-0.02\textwidth}
    \includegraphics[width=0.26\textwidth]{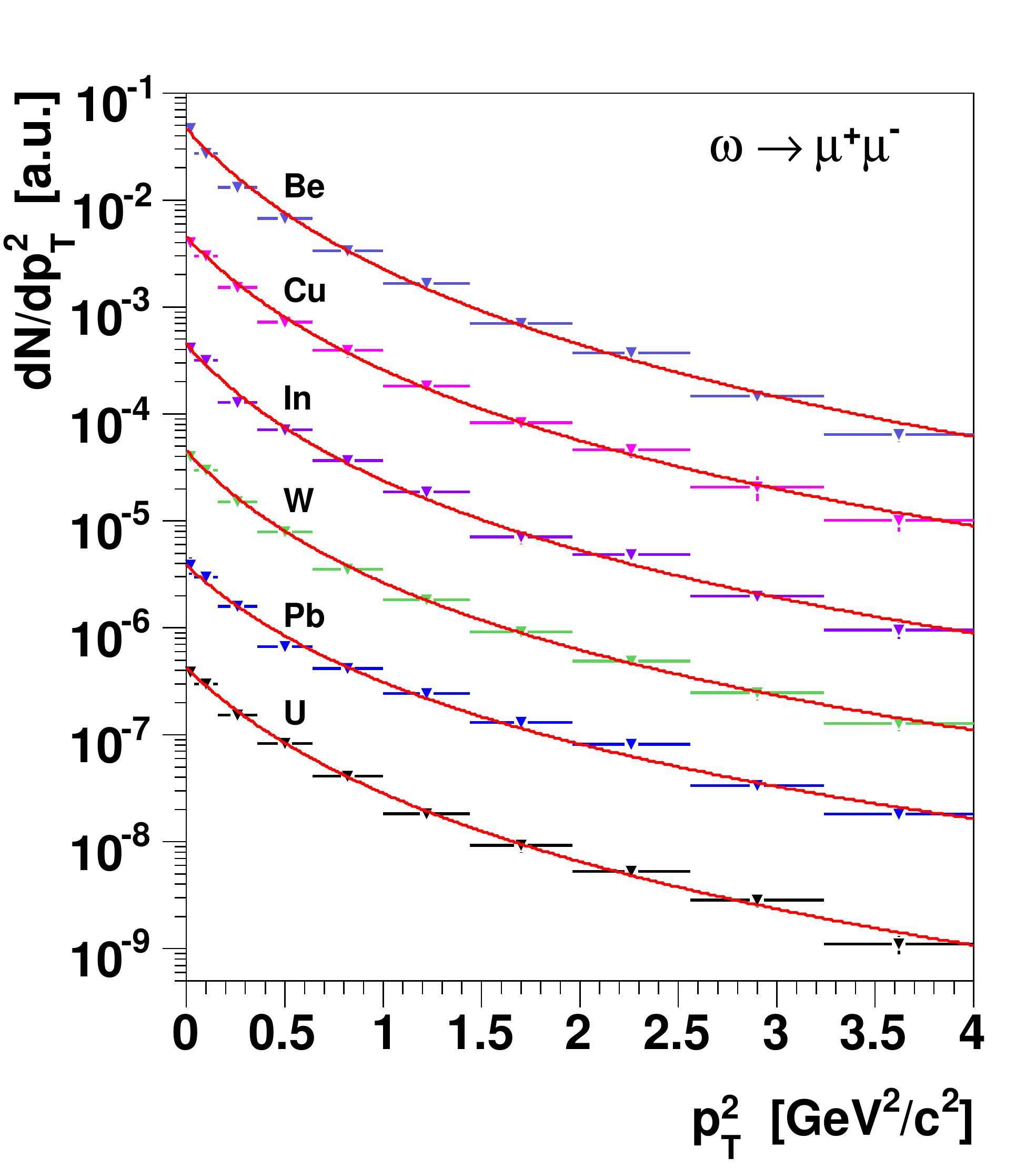}\hspace*{-0.02\textwidth} 
    \includegraphics[width=0.26\textwidth]{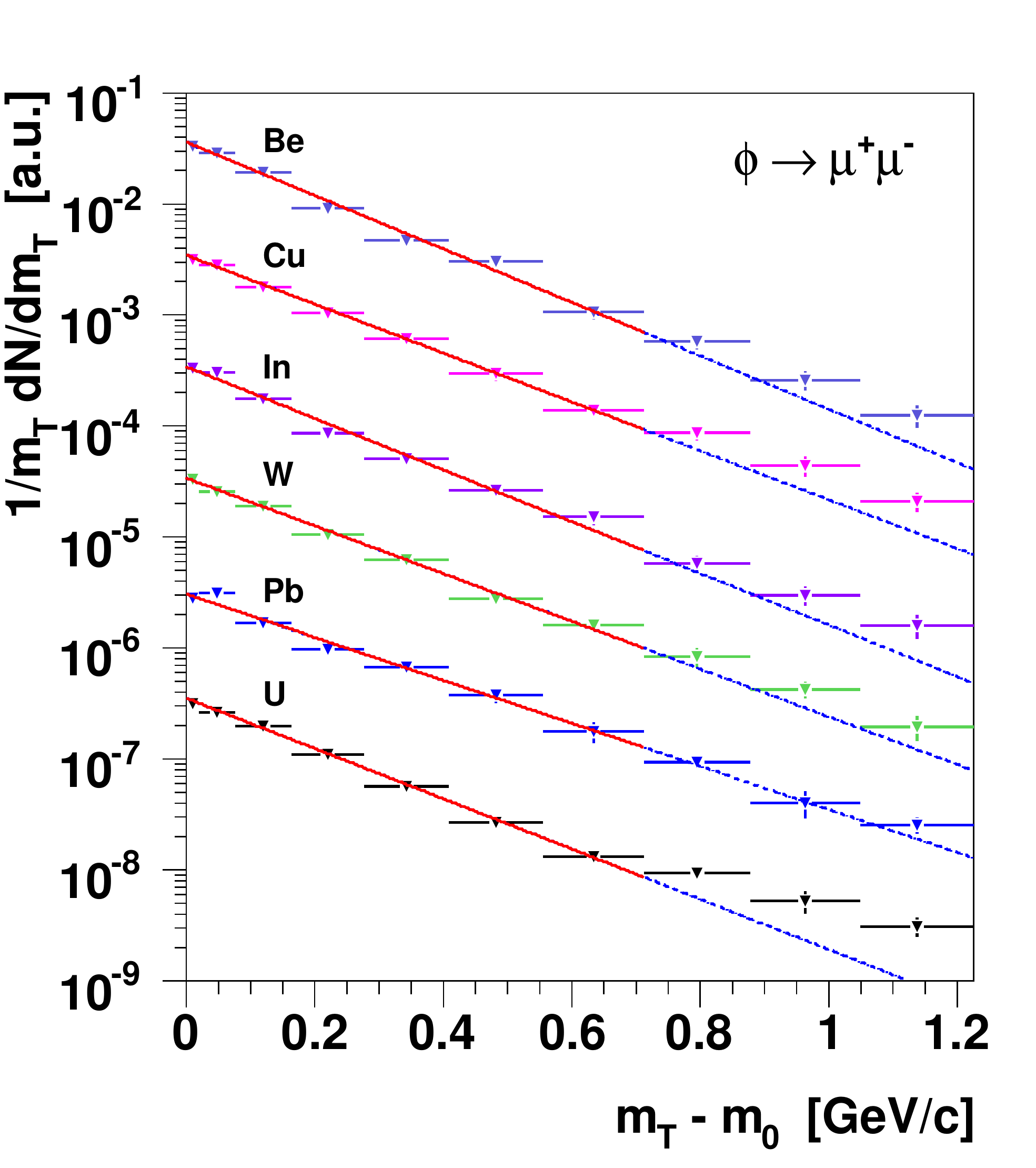}\hspace*{-0.02\textwidth}
    \includegraphics[width=0.26\textwidth]{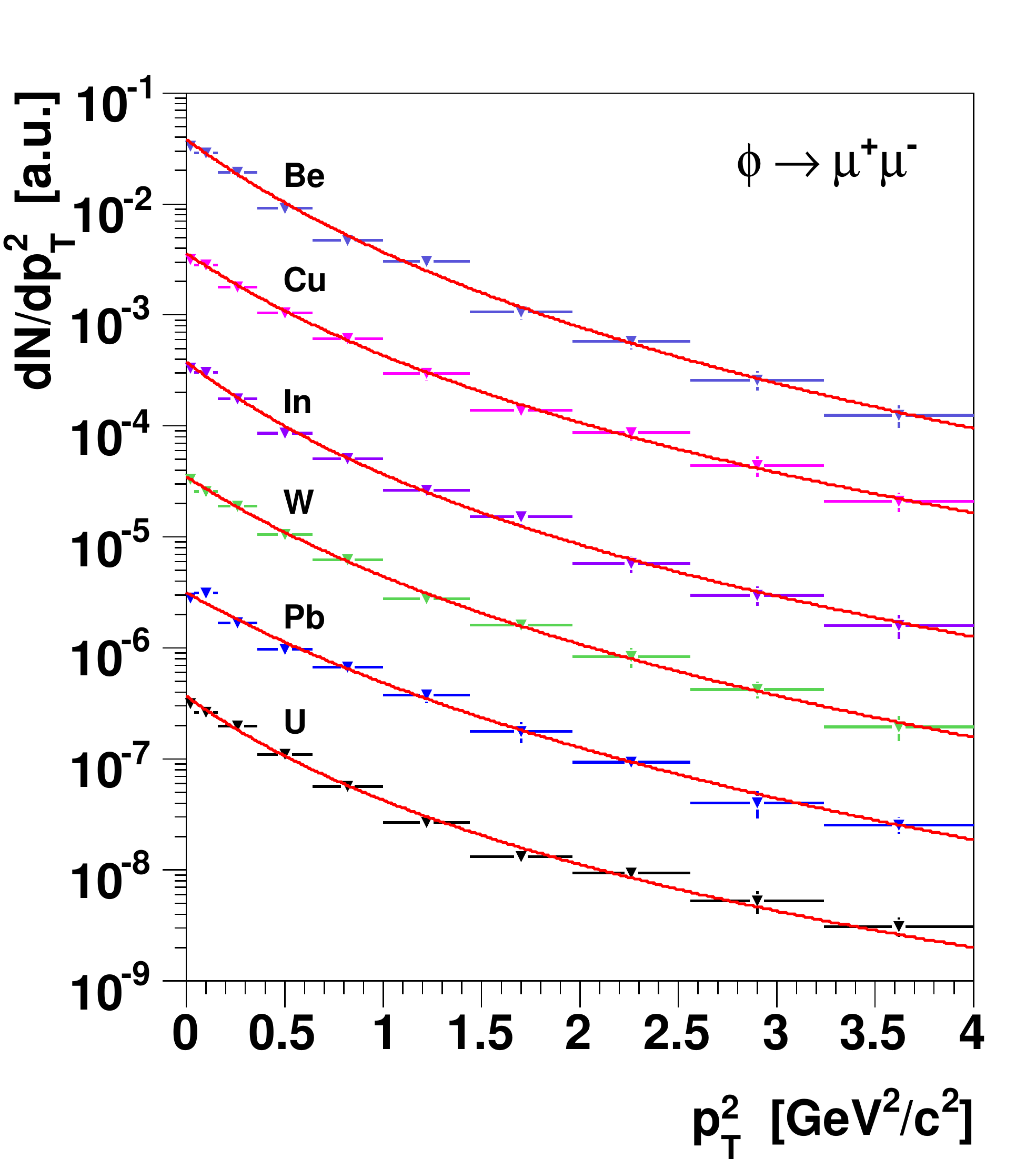}\hspace*{-0.06\textwidth}
  \end{center}
\vspace{-0.6cm}
\caption[\textwidth]{Fits on the acceptance-corrected $\mt$ and $\pt^2$ spectra}
\label{fig:pt_spectra}
\end{figure}

%\section{Transverse Momentum Spectra for the $\omega$ and $\phi$ Mesons}
\noindent In order to study the $\pt$ spectra, the data sample was divided, target by target, into $\pt$ bins of 200 MeV/$c$. For each target and $\pt$ bin, a fit is then performed on the raw mass spectrum with the superposition of the expected sources. In this way, a raw $\pt$ spectrum is extracted for the $\omega$ and $\phi$ mesons, which is then corrected for the acceptance and efficiency as a function of $\pt$, as given by the MC simulations. The $\omega$ and $\phi$ $\pt$ spectra resulting after the correction for the acceptance $\times$ efficiency are shown in \figurename~\ref{fig:pt_spectra} as a function of the transverse mass $\mt$ and $\pt^2$. The $\mt$ spectra have been compared to a thermal-like function $ \d N/(\pt\, \d \pt) = \d N/(\mt \,\d \mt) \propto \exp ( -\mt/T)~$. One can immediately appreciate how the thermal hypothesis clearly fails in describing the whole spectra up to 2~GeV/$c$, a systematical deviation from the pure exponential trend being clearly  visible for high $\mt$. On the contrary, the power-law function $\d N/\d \pt^2 \propto \left( 1 + \pt^2/p_0^2 \right)^{-\beta}$, a standard form for a mixture of soft and hard processes, is adequate to describe the $\d N /\d \pt^2$ distribution in the whole range considered in the present analysis. No significant trend for the mean value $\langle \pt \rangle$ as function of~A can be identified. The average $\pt$ integrated over the different nuclear targets are: $\langle \pt \rangle_\omega = 0.61 \pm 0.03~\mathrm{(stat.)} \pm 0.03~\mathrm{(syst.)}~\mathrm{GeV}/c$, $\langle \pt \rangle_\phi = 0.70 \pm 0.04~\mathrm{(stat.)} \pm 0.02~\mathrm{(syst.)}~\mathrm{GeV}/c$. These results can be compared with the NA27 measurements in p-p collisions at $\sqrt{s} = 27.5$~GeV~\cite{Verbeure:1991tv}: $\langle \pt \rangle_\omega^\mathrm{NA27} = (0.591 \pm 0.021)~\mathrm{GeV}/c$ and $\langle \pt \rangle_\phi^\mathrm{NA27} = (0.513 \pm 0.030)~\mathrm{GeV}/c$. As one can see, the NA27 value for $\langle \pt \rangle$ for the $\omega$ agrees with the one obtained in the present analysis. On the other hand, for the $\langle \pt \rangle$ of the $\phi$, there is a disagreement by more than 4 (statistical) standard deviations. Furthermore, the NA60 results clearly indicate that $\langle \pt \rangle_\phi > \langle \pt \rangle_\omega$, while the NA27 results indicate that $\langle \pt \rangle_\phi < \langle \pt \rangle_\omega$. Concerning the $\phi$ meson, HERA-B measured $\phi$ production in p-A at $\sqrt{s} = 41.6$~GeV~\cite{Abt:2006wt}. The $\pt$ distributions measured by HERA-B are well described by the power-law function cited above, with $\langle \pt \rangle$ values fully compatible with the NA60 ones and no definite trend as a function of~A.

\begin{figure}[htbp] 
   \begin{center} 
   \includegraphics[width=0.36\textwidth]{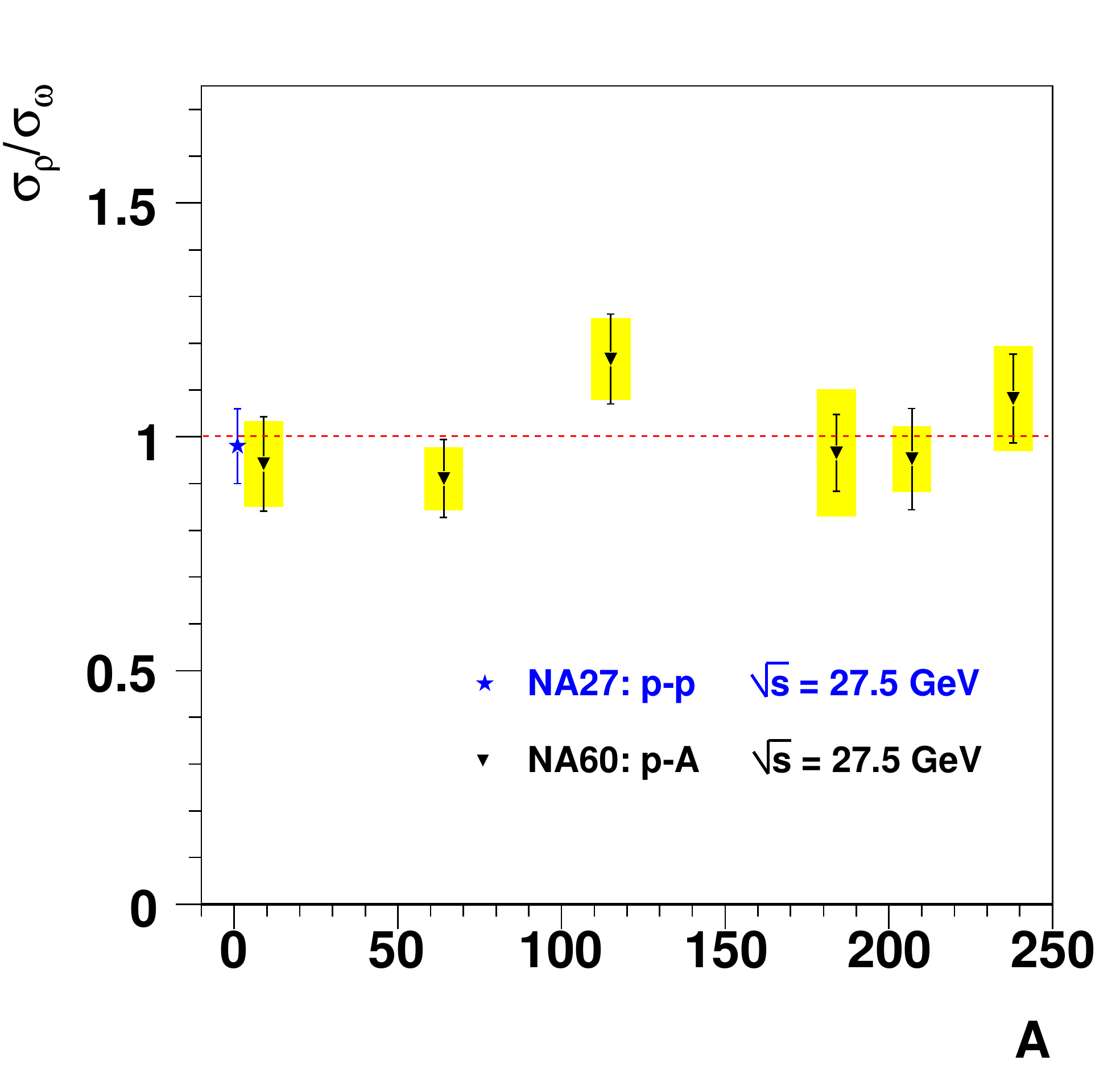} \hspace{0.08\textwidth} 
   \includegraphics[width=0.36\textwidth]{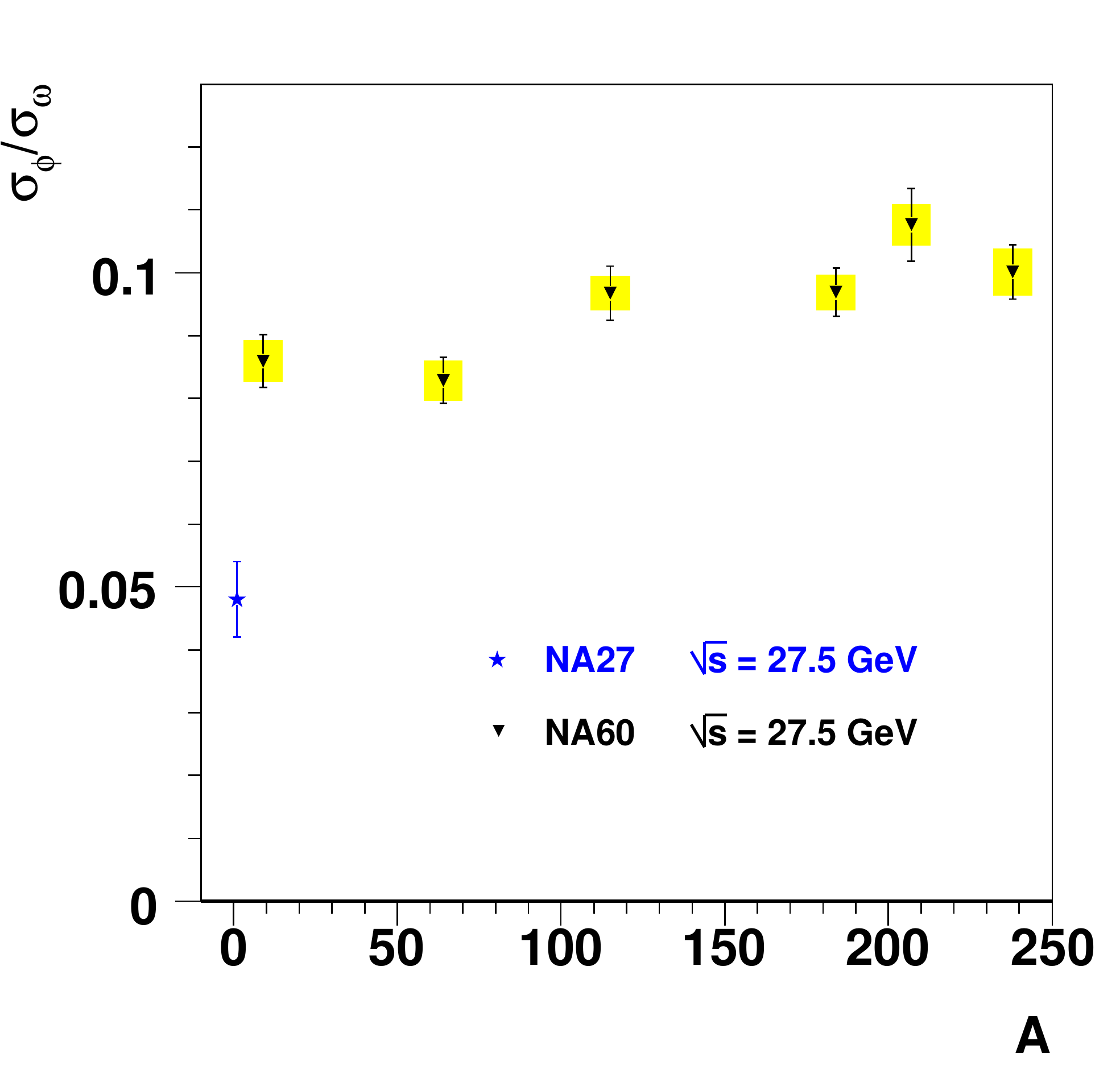}
    \end{center} 
\vspace{-0.7cm}
\caption[\textwidth]{Production cross section ratios $\sigma_\rho/\sigma_\omega$ and $\sigma_\phi/\sigma_\omega$ in full phase space. Error bars and shadowed boxes account for statistical and systematic uncertainties, respectively.}
\label{fig:particleRatiosRhoPhi}
\end{figure}

%\section{Production of $\eta$, $\rho$, $\omega$ and $\phi$ Mesons}
~\\
\noindent Fitting the raw mass spectra target by target allows us to study the nuclear dependence of particle production. We first consider the particle ratios integrated over $\pt$, normalizing the cross sections to the one of the $\omega$ meson as shown in~\figurename~\ref{fig:particleRatiosRhoPhi}. The $\sigma_\rho/\sigma_\omega$ ratio appears to be flat with~A and consistent with $\sigma_\rho/\sigma_\omega = 1$. The ratio averaged over~A $-$ indicated by an horizontal line in the figure $-$ is $\sigma_\rho / \sigma_\omega = 1.00 \pm 0.04~\mathrm{(stat.)} \pm 0.04~\mathrm{(syst.)}$. This is in agreement with the ratio $\sigma_\rho/\sigma_\omega=0.98\pm0.08$ measured in p-p collisions at $\sqrt{s}=27.5$~GeV by the NA27 experiment~\cite{Verbeure:1991tv}. A rather different conclusion can be drawn, on the contrary, for the $\sigma_\phi/\sigma_\omega$ ratio, for which the p-p measurement by NA27~\cite{Verbeure:1991tv} is also shown: the trend of the data points indicates that a very significant strangeness enhancement is already observed going from p-p to p-A collisions, with a further increase as a function of A.

\noindent The nuclear dependence of the production cross sections has been studied, for each meson, normalizing the measured yields to the one of the lightest target, namely the Be one. The resulting trend is satisfactorily described by the power-law $\sigma_\mathrm{pA} \propto \mathrm{A}^\alpha$, leading to the following estimation for the $\alpha$ parameter for the $\omega$ and $\phi$ mesons, integrated over $\pt$: $\alpha_\omega = 0.841 \pm 0.014~\mathrm{(stat.)} \pm 0.030~\mathrm{(syst.)}$, $\alpha_\phi = 0.906 \pm 0.011~\mathrm{(stat.)} \pm 0.025~\mathrm{(syst.)}$. A possible bias affecting these results comes from the fact that the Be target may start to deviate from the power law above, because of its low mass number making it more similar to an incoherent superposition of single nucleons, rather than to a nuclear system having collective properties. Indeed, if one removes the Be target from the analysis, the values for the $\alpha$ parameters shift to $\alpha_\omega = 0.801 \pm 0.037~\mathrm{(stat.)} \pm 0.041~\mathrm{(syst.)}$ and $\alpha_\phi = 0.963 \pm 0.026~\mathrm{(stat.)} \pm 0.017~\mathrm{(syst.)}$, the $\omega$ moving towards a softer nuclear dependence, a harder one being found, on the contrary, for the $\phi$.

\begin{figure}[htbp] 
   \begin{center}
    \includegraphics[width=0.32\textwidth]{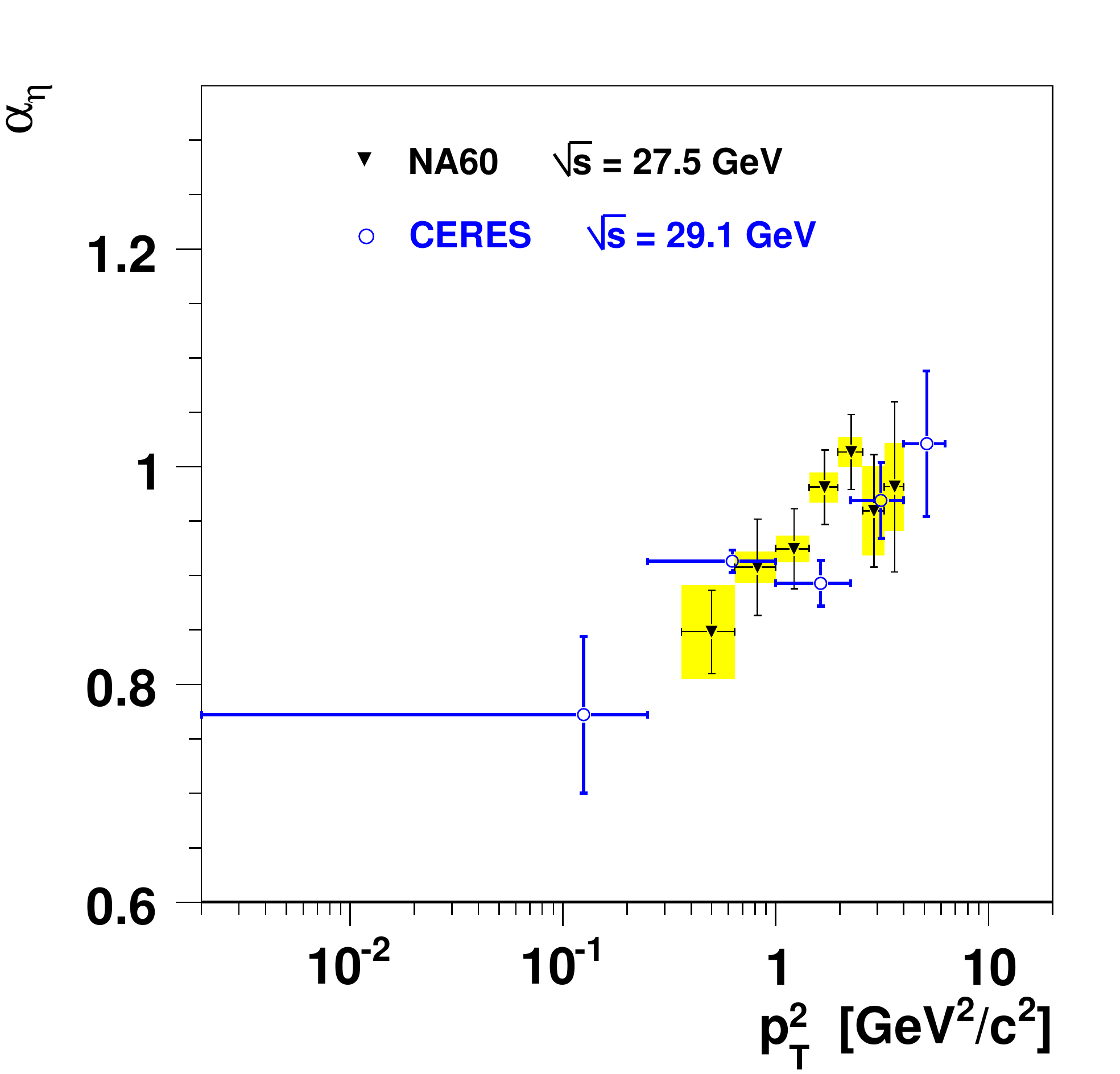}
    \includegraphics[width=0.32\textwidth]{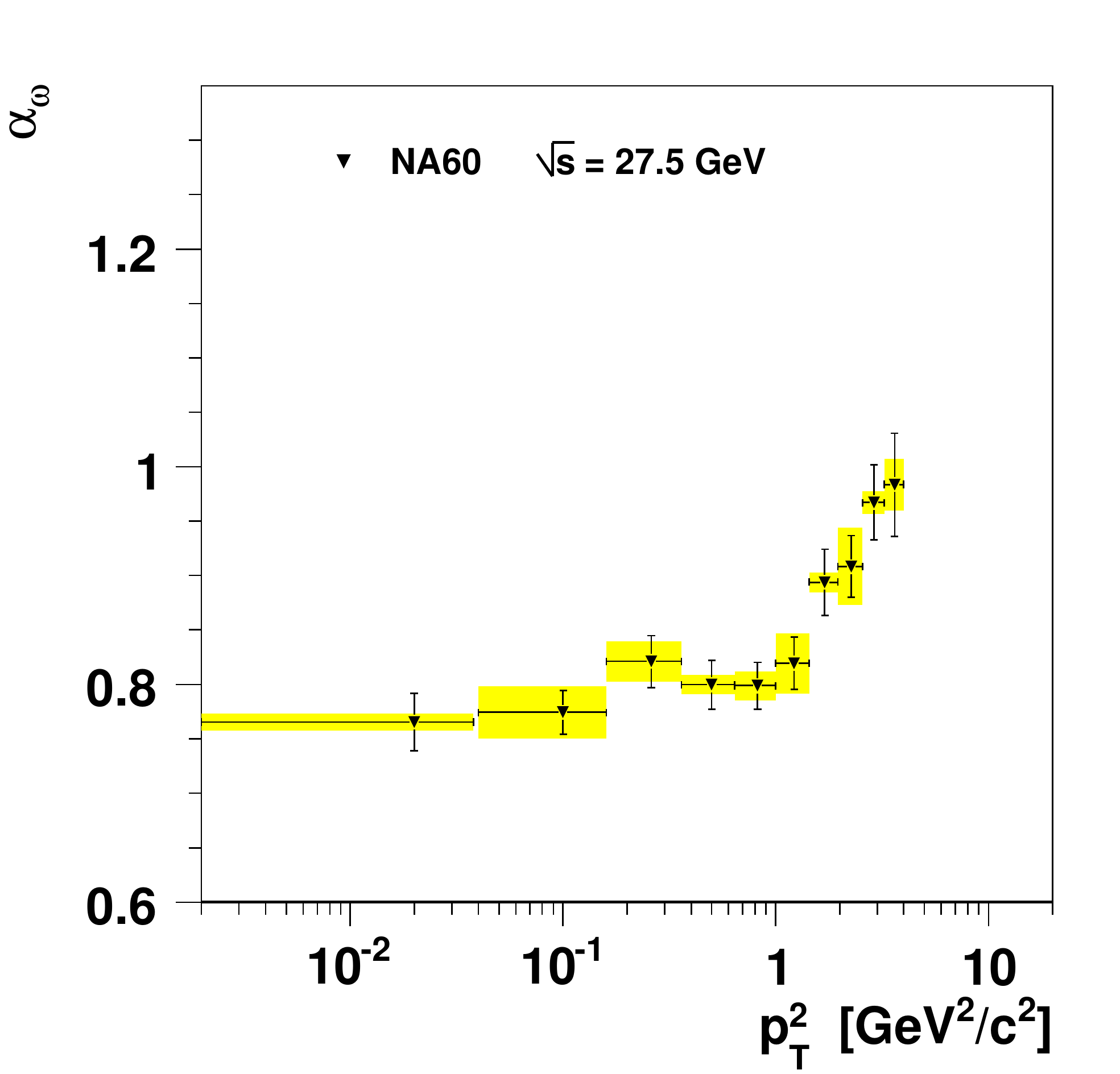} 
    \includegraphics[width=0.32\textwidth]{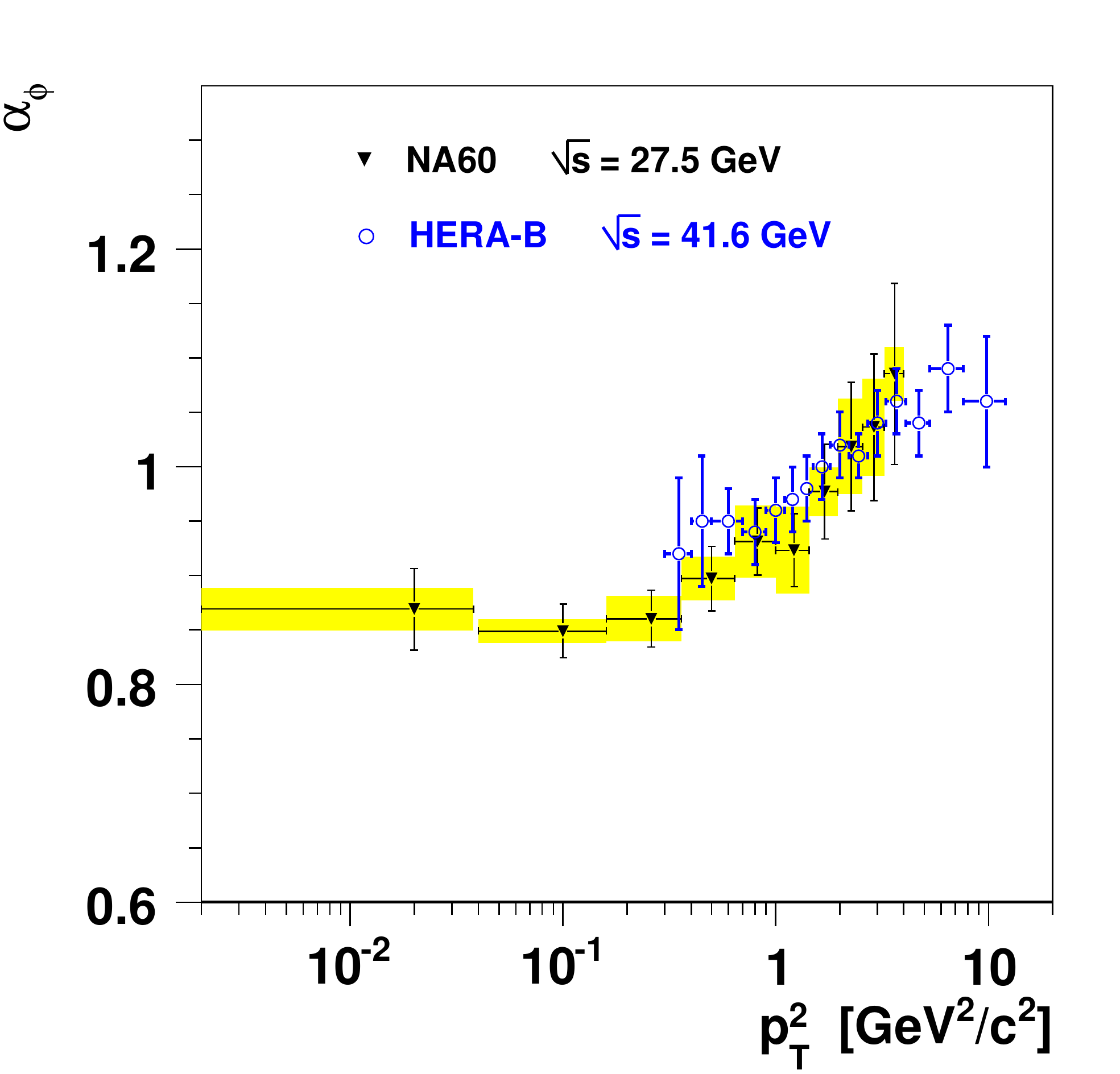}
    \end{center} 
\vspace{-0.5cm}
\caption[\textwidth]{$\pt$ dependence of the $\alpha$ parameter for the $\eta$, $\omega$ and $\phi$ mesons.}
\label{fig:alphaVsPt}
\end{figure}

\noindent The available data sample also allowed us to investigate the $\pt$ dependence of the $\alpha$ parameters. This study has also been performed for the $\eta$ meson, for which the $\pt$ coverage starts from $0.6$~GeV/$c$. The results, reported in \figurename~\ref{fig:alphaVsPt}, clearly indicate an increase of the $\alpha$ parameters as a function of $\pt$, which can be related to the so-called ``Cronin effect'' originally observed by Cronin~\emph{et al.}~for charged kaons~\cite{Kluberg:1977bm}. The same observation is reported for the $\phi$ meson by the HERA-B Collaboration~\cite{Abt:2006wt}, whose points are in remarkable agreement with the NA60 ones. A similar comparison can be established for the $\eta$ meson, thanks to the data published by the CERES-TAPS Collaboration~\cite{Agakishiev:1998mw}: here, too, a good agreement is observed between the two sets of data points. No comparison was possible for the $\omega$, due to the lack of available measurements.

%\section*{References}

\end{document}